\documentclass[aps,pra,twocolumn,superscriptaddress,preprintnumbers,nofootinbib]{revtex4-2}
\usepackage[utf8]{inputenc}
\usepackage{graphicx}
\usepackage{amsmath}
\usepackage{amsthm}
\usepackage{bm}
\usepackage{layout}
\usepackage{float}
\usepackage{amsfonts}
\usepackage{amssymb}%
\usepackage{array}%
\usepackage[margin=0.75in]{geometry}
\usepackage{color}
\usepackage{soul}
\usepackage{mathtools}
\usepackage[colorlinks=true,citecolor=blue]{hyperref}
\theoremstyle{definition}

\theoremstyle{theorem}

\usepackage{physics}

\usepackage{ulem}

\renewcommand{\emph}[1]{\textit{#1}}

\newcommand{\red}[1]{{\color{red} #1}}

\newcounter{para}

\makeatletter
\newcommand*\bigcdot{\mathpalette\bigcdot@{.5}}
\newcommand*\bigcdot@[2]{\mathbin{\vcenter{\hbox{\scalebox{#2}{$\m@th#1\bullet$}}}}}

\newcommand{\llangle}[1][]{\savebox{\@brx}{\(\m@th{#1\langle}\)}%
  \mathopen{\copy\@brx\kern-0.5\wd\@brx\usebox{\@brx}}}
\newcommand{\rrangle}[1][]{\savebox{\@brx}{\(\m@th{#1\rangle}\)}%
  \mathclose{\copy\@brx\kern-0.5\wd\@brx\usebox{\@brx}}}

\makeatother

\usepackage{array}
\newcolumntype{L}{>{$}l<{$}} 
\newcolumntype{C}{>{$}c<{$}} 
\newcolumntype{R}{>{$}r<{$}} 

\usepackage{xcolor}

\usepackage{mathtools}

\usepackage{physics}
\usepackage{bbm}
\usepackage{booktabs}

\newtheorem*{theorem*}{Theorem}
\newtheorem*{lemma*}{Lemma}

\usepackage{comment}
\usepackage{ragged2e}
\usepackage{atbegshi,picture}
\usepackage{array}
\usepackage{hyperref}
\usepackage{tocloft}
\usepackage{placeins}
\usepackage{tcolorbox}
\usepackage{capt-of}

\newcolumntype{M}[1]{>{\centering\arraybackslash}m{#1}}
\usepackage{soul}
 \makeatletter
\renewcommand{\@cftmaketoctitle}{} 
\makeatother
 
\usepackage{newfloat}
\DeclareFloatingEnvironment{boxes} 
\newcommand{\boxref}[1]{\hyperref[{#1}]{Box~\ref*{#1}}}

\begin{document}
\newcommand{\Caltech}{California Institute of Technology, Pasadena, CA, USA}
\newcommand{\MIT}{
Center for Theoretical Physics --- a Leinweber Institute, 
Massachusetts Institute of Technology, Cambridge, MA, USA}
\newcommand{\WalterBurke}{Walter Burke Institute for Theoretical Physics, California Institute of Technology, Pasadena, CA, USA}
\newcommand{\UTAustin}{
Department of Physics, The University of Texas at Austin, Austin, Texas 78712, USA}
\newcommand{\NUS}{ 
Department of Physics, 
National University of Singapore, Singapore 117551}
\newcommand{\CQT}{ 
Centre for Quantum Technologies, National University of Singapore, Singapore 117543}

\newcommand{\TKcite}{{\color{red}[CITE]}}
\newcommand{\nocontentsline}[3]{}
\let\origcontentsline\addcontentsline
\newcommand\stoptoc{\let\addcontentsline\nocontentsline}
\newcommand\resumetoc{\let\addcontentsline\origcontentsline}

\title{
Deep thermalization and Hilbert space ergodicity
}

\author{Daniel~K.~Mark}
\affiliation{\MIT}
\author{Manuel~Endres}
\affiliation{\Caltech}
\author{Matteo Ippoliti}
\altaffiliation{Corresponding author: \href{mailto:ippoliti@utexas.edu}{ippoliti@utexas.edu}}
\affiliation{\UTAustin}

\author{Wen Wei Ho}
\altaffiliation{Corresponding author: \href{mailto:wenweiho@nus.edu.sg}{wenweiho@nus.edu.sg}}
\affiliation{\NUS}
\affiliation{\CQT}

\author{Soonwon~Choi}
\altaffiliation{Corresponding author: \href{mailto:soonwon@mit.edu}{soonwon@mit.edu}}
\affiliation{\MIT}

\begin{abstract}
Recent advances in quantum simulation have enabled the discovery of novel forms of universality in quantum many-body dynamics. 
In this Review, we discuss \textit{deep thermalization} and \textit{Hilbert space ergodicity}--two recently discovered phenomena 
characterized by the emergence of distributions of quantum states that are ``maximally random" in a precise sense. 
They offer a new perspective on how irreversible statistical mechanics arises from reversible unitary quantum dynamics, 
going beyond conventional theories of quantum thermalization and equilibration.
We review a unifying framework, rooted in quantum information theory and principles of maximum entropy, that explains the different forms of ergodicity which arise under various physical constraints.
We discuss open questions and active research directions, including generalizations beyond ensembles of pure quantum states, phase transitions in deep thermalization tied to deeper forms of ergodicity-breaking, 
connections to broader topics in quantum thermalization and ergodicity, and applications in quantum information science, such as for benchmarking or tomography. 
\end{abstract}

\maketitle

\noindent\textbf{Introduction.} 
Ergodicity is a foundational concept underlying statistical approaches to many-body physics, and provides a microscopic basis for the emergence of thermodynamic behavior.
In classical systems, ergodicity is formalized by Birkhoff's theorem, which relates the time trajectory of a single system to the distribution of an ensemble of systems over phase space adhering to maximum entropy principles~\cite{brin2002introduction}.
In quantum systems, where dynamics takes place in Hilbert space instead of classical phase space, it is a deep and long-standing question how these   notions of ergodicity and maximum entropy can be generalized~\cite{ho2018ergodicity}.  
Quantum thermalization offers one powerful answer, 
based on the idea of the quantum many-body system acting as its own ``heat bath'': 
due to the generation of entanglement between different constituents, local subsystems can approach 
mixed thermal states, even when the global system remains pure. 
Understanding the precise conditions under which thermalization occurs, and the mechanisms by which it can fail, is an active area of research~\cite{abanin2019colloquium,moudgalya2022quantum,gogolin2011absence}. 

\begin{figure*}[tpb!]
    \centering
    \includegraphics[width=\linewidth]{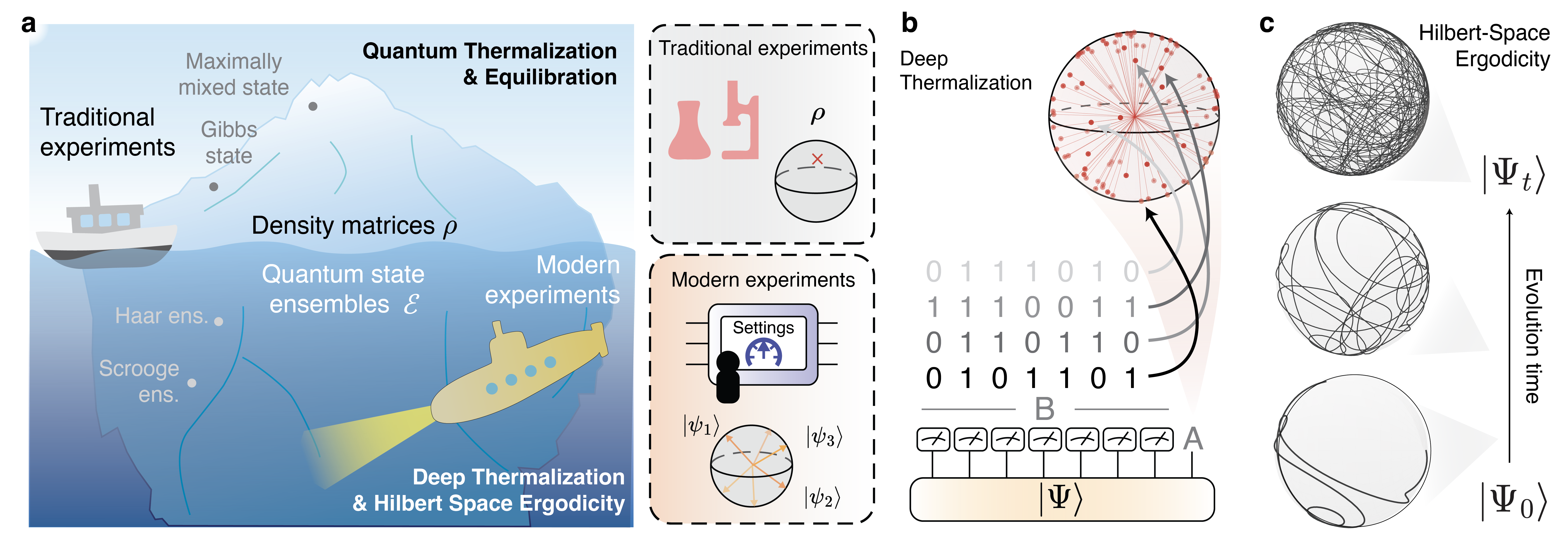}
    \caption{\textbf{New forms of quantum ergodicity.} \textbf{a.} Quantum thermalization as probed by traditional experiments is only the tip of the iceberg 
    of ergodicity in quantum many-body systems. In contrast, modern experiments can probe hidden, deeper forms of universality. In this Review, we discuss two phenomena discovered by these new measurement capabilities: \textit{deep thermalization} and \textit{Hilbert space ergodicity}. These describe universal features of {\it quantum state ensembles} 
    generated through measurements of a bath or through time evolution, that go beyond the {\it density matrices} probed by traditional experiments (right column, adapted from Ref.~\cite{mark2024maximum}).
    \textbf{b.} Deep thermalization describes the emergence of universal randomness in the \textit{projected ensemble} $\{p(z_B), |\Psi_A(z_B)\rangle\}$, the collection of local conditional states  on subsystem $A$ tied to measurement outcomes $z_B$ on the complement $B$ and their associated Born probabilities. 
    As an example, 
    the projected ensemble may approach the so-called \textit{Haar ensemble}, the uniform distribution over Hilbert space (red points on Bloch sphere)~\cite{choi2023preparing,cotler2023emergent,ho2021exact}.
    \textbf{c.} Hilbert space ergodicity (HSE) studies the \textit{temporal ensemble}: the trajectory of a quantum state in dynamics. HSE is the phenomenon whereby the state eventually explores its available Hilbert space uniformly, subject to the presence of constraints such as energy conservation.
    }
\label{fig:overview}
\end{figure*}

The conventional framework of quantum thermalization is 
tailored to traditional solid-state and condensed-matter systems, where experimentally accessible quantities are primarily expectation values of local or few-body observables. However, rapid experimental advances in controlling systems of many interacting quantum particles have made it possible to probe many-body systems at a level of detail far beyond such observables, motivating questions that extend beyond the traditional scope of thermalization (Fig.~\ref{fig:overview}).
Indeed, modern quantum simulation platforms such as arrays of neutral atoms~\cite{menssen2026strategic}, trapped ions~\cite{moses2023racetrack}, and superconducting quantum processors~\cite{arute2019quantum} offer unprecedented control and the ability to perform highly resolved, nonlocal measurements, ranging from configurational snapshots of every particle in the system to information-theoretic properties such as entanglement entropy~\cite{kaufman2016quantum} and out-of-time-ordered correlators~\cite{google2025constructive}. 
This newfound ability to access fine-grained and nonlocal information raises a natural question: what should quantum ergodicity and chaos mean when one probes not only expectation values, but the full statistical structure of quantum states? 

{\it Deep thermalization} and {\it Hilbert-space ergodicity} (HSE) are two recently developed answers to this question. Rather than characterizing only the average behavior of conventional observables, these frameworks probe higher-order statistical information, such as fluctuations encoded in system-bath correlations or  fluctuations over time. The  natural language for describing this richer physics is that of \textit{quantum state ensembles}: probability distributions 
over Hilbert space, which contain information beyond the often-studied density matrices. 
Deep thermalization and HSE represent the discovery of  universality in these new settings: quantum state ensembles arising from sampling the time evolution of a system (HSE) or the state of a bath (deep thermalization) have been found to be {\it maximally random} in generic quantum many-body systems, up to physical constraints such as conservation laws. This idea is
made precise by maximum entropy principles rooted in quantum information theory~\cite{holevo1998capacity,jozsa1994lower}.
Deep thermalization and HSE therefore represent exciting new developments in our fundamental understanding of ergodicity and chaos in quantum many-body systems.
Moreover, the emergent randomness characterizing these phenomena has promising potential applications in quantum information science: it has been proposed to be a resource for quantum state benchmarking, tomography, and certification~\cite{choi2023preparing,mark2022benchmarking,tran2023measuring,huang2025certifying}. 

The goal of this Review is to survey the latest progress in the understanding of deep thermalization and Hilbert-space ergodicity, and to identify important open questions and promising directions for future work. We introduce the theoretical framework underlying these phenomena, review their experimental evidence, discuss connections to broader notions of quantum ergodicity and chaos, and highlight applications in quantum information science. 

\noindent\textbf{Ensembles of quantum states.}  
The key objects of study in deep thermalization and HSE are  quantum state ensembles: 
\begin{equation}
    \mathcal{E} = \{ p_j, |\Psi_j\rangle \},
\end{equation}
collections of wavefunctions $|\Psi_j\rangle$ weighted by probabilities $p_j$, where the labels $j$ may be discrete or continuous (in which case the probabilities $p_j$ correspond to a continuous probability distribution $p(\Psi) d\Psi$ over Hilbert space).
Quantum state ensembles have long been studied 
as theoretical objects in statistical mechanics~\cite{park1977rigorous,brody1999geometrization,willson2026maximum,anza2026nonequilibrium}, as well as
in quantum information theory, for example in the context of information transmission~\cite{jozsa1994lower,holevo1998capacity}, 
where a state $|\Psi_j\rangle$ corresponds to  a quantum encoding of a classical message $j$ (or `codeword') to be transmitted.

The novelty of deep thermalization and HSE lies in the fact that they concern state ensembles which arise in 
{\it physical} settings, 
such as the quantum states induced by interactions with an environment or bath, 
in which case the  label $j$ carries the meaning of a particular configuration  $z$ of the bath; or those arising from time evolution under a chaotic Hamiltonian or circuit, in which case $j$ has the meaning of the particular time $t$ the system is evolved to.
Traditionally, this type of fine-grained information has not been readily accessible in the laboratory: e.g., solid-state or condensed matter experiments typically probe spatially or temporally {\it averaged} properties of a physical system, encapsulated by the density matrix $\rho = \sum_j p_j |\Psi_j\rangle \langle \Psi_j|$. 
However, the advent of programmable, high-fidelity quantum simulators with unprecedented measurement capabilities has changed this premise, 
prompting the following questions:

{\it
Do quantum state ensembles arising in physical many-body settings exhibit universal structures  (i.e., independent of precise microscopic details) that go beyond the density matrix? If so, how can we characterize them, and what physical principles dictate their emergence?
}

Indeed, a quantum state ensemble $\mathcal{E}$ contains information not captured by its density matrix $\rho$, which is only the mean state of the ensemble. Instead, the former encodes {\it fluctuations} about the latter. Notably, many state ensembles can give rise to the same density matrix--in quantum information parlance, they correspond to different ``unravelings'' of $\rho$. Fluctuations constitute a unique fingerprint identifying each ensemble, and serve as a measurable diagnostic for distinguishing between them.

Concretely,  fluctuations of an ensemble $\mathcal{E}$ can be systematically quantified through its  statistical moments. 
The $k$-th moment operator $\rho^{(k)}_\mathcal{E}$ is defined by averaging $k = 1,2,\dots$ identical copies of each state in $\mathcal{E}$, 
\begin{equation}
\rho^{(k)}_{\mathcal{E}} := \mathbb{E}_{\Psi\sim \mathcal{E}} \left[|\Psi\rangle \langle \Psi|^{\otimes k} \right],
\end{equation}
where $\displaystyle \mathbb{E}$ denotes ensemble-averaging. Note that  $\rho^{(k)}_{\mathcal{E}}$ is a bona fide density matrix 
defined on the $k$-copy space 
$\mathcal{H}^{\otimes k}$: it has unit trace and is positive semi-definite. The first moment $\rho^{(1)}_\mathcal{E}$ is the familiar density matrix $\rho$, which encodes ensemble-averaged expectation values of observables. Meanwhile, the second moment $\rho^{(2)}_\mathcal{E}$  determines their variance over the ensemble.  Specifically, the (co)variance of any two observables $O_1, O_2$ over $\mathcal{E}$ is 
\begin{align}
\text{cov}_\mathcal{E}(O_1,O_2) := \sum_j p_j \langle \psi_j |O_1|\psi_j\rangle\!\langle \psi_j |O_2|\psi_j\rangle & \\
- 
\big(\sum_j p_j \langle \psi_j |O_1 |\psi_j\rangle\big) 
\big(\sum_j p_j \langle \psi_j |O_2 |\psi_j\rangle\big)&, \nonumber
\end{align}
which can be compactly expressed in terms of the moment operators as $\text{tr}(\rho^{(2)}_\mathcal{E}O_1\otimes O_2) - \text{tr}(\rho^{(1)}_\mathcal{E} O_1)\text{tr}(\rho^{(1)}_\mathcal{E}  O_2)$. 
To compare between two ensembles $\mathcal{E}$ and $\mathcal{E}'$, a standard metric is the trace distance of their $k$-th moment operators 
\begin{equation}
    \Delta^{(k)}(\mathcal{E},\mathcal{E}') \equiv \frac{1}{2} \left\Vert \rho^{(k)}_\mathcal{E} - \rho^{(k)}_{\mathcal{E}'} \right\Vert_*,
    \label{eq:k_trace_dist}
\end{equation}
where $\Vert \cdot \Vert_*$ is the trace norm. 
The vanishing of $\Delta^{(k)}$ 
implies the statistical indistinguishability of the two ensembles using any protocol involving $k$ copies of the states, with close connection to the notion of state designs (Box~\red{1}).


\noindent\textbf{Deep thermalization and Hilbert space ergodicity.} 
Deep thermalization and HSE describe the emergence of  maximal randomness across all statistical moments of 
two key quantum state ensembles in physics: the projected ensemble (arising from system-bath interactions) and the temporal ensemble (emerging over time under chaotic dynamics). They represent a refined generalization of standard quantum thermalization, which prescribes only the structure of the first moments--namely, that the reduced density matrices of small subsystems~\cite{pilatowsky2025quantum} and the diagonal ensemble~\cite{gogolin2011absence} respectively converge to thermal Gibbs states.

Projected ensembles are defined as follows. 
Consider a single quantum many-body state $|\Psi\rangle_{AB}$, which could be a state derived from late-time dynamics or an energy eigenstate of a Hamiltonian etc., 
and a bipartition of the system into a small subsystem ($A$) and its large complement  ($B$), which is treated as the subsystem's ``bath.''
Assume now that one performs measurements of the subsystem $B$ in a fixed local basis $\{ |z_B\rangle\}$. Then,  outcome $z_B$ is observed with probability $p(z_B)$, upon which the global state collapses into 
$|\Psi_A(z_B)\rangle\otimes|z_B\rangle$. 
We identify $|\Psi_A(z_B)\rangle$ as the \textit{projected state} on A, and call the collection of all projected states the \textit{projected ensemble}~\cite{choi2023preparing,cotler2023emergent}:
\begin{align}
    \mathcal{E}_\text{Proj.} & := \{p(z_B), |\Psi_A(z_B)\rangle \}_{z_B}~. \label{eq:proj_ens}
\end{align}
Note that there are exponentially many (in the size $|B|$) possible measurement outcomes $z_B$ and hence projected states; ignoring these outcomes returns the familiar reduced density matrix on $A$, $\rho_A = \text{tr}_B|\Psi\rangle\langle\Psi| = \rho^{(1)}_{\mathcal{E}_\text{Proj.}}$, which captures strictly local behavior of the system. 
In contrast, the full distribution of states given by the projected ensemble $\mathcal{E}_\text{Proj.}$ captures detailed local subsystem behavior {\it conditioned} on information about the bath (specifically, on a classical configuration $z_B$ the bath can be found in).

Temporal ensembles are instead generated through unitary quantum dynamics, for example by a given Hamiltonian $H$ or circuit.
The \textit{temporal ensemble} 
\begin{align}
    \mathcal{E}_\text{Temp.} & := \{|\Psi_t\rangle\}_{t \in [0,T]}
    \label{eq:temp_ens}
\end{align} 
is defined as the collection of states visited in the trajectory of a single initial state $|\Psi_0\rangle$ over the time interval $[0,T]$.
Ignoring the ensemble label $t$ 
amounts to probing the system's time-averaged behavior, captured by the so-called  diagonal ensemble~\cite{gogolin2011absence} $\rho_d \equiv \mathbb{E}_t[|\Psi(t)\rangle\langle\Psi(t)|]$. In contrast, the full distribution of states specified by the temporal ensemble $\mathcal{E}_\text{Temp.}$ captures the detailed structure of fluctuations over time of a quantum state as it traverses the  Hilbert space.

\begin{tcolorbox}[float,floatplacement=t,label={box2:state_designs}]
{\textbf{Box~1: Haar ensembles and quantum state designs}}

\vspace{2mm}
\includegraphics[width=\textwidth]{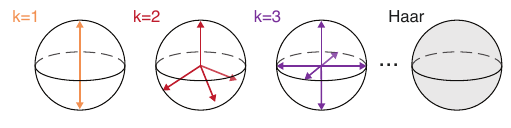}
\vspace{-6mm}
\captionof{figure}{\textbf{Quantum state designs}. Adapted from Ref.~\cite{cotler2023emergent}.}

Quantum state $k$-designs are ensembles of quantum states that are indistinguishable from the Haar random ensemble up to the $k$th statistical moment~\cite{ambainis2007quantum}.
They are of particular interest in quantum information, since these moments have explicit analytic expressions~\cite{mele2024interoduction} which enable applications including cryptography~\cite{hayden2004randomization}, benchmarking~\cite{arute2019quantum}, and tomography~\cite{elben2023randomized}. 
State designs are useful because they can be constructed with considerably fewer resources than the Haar ensemble while still possessing the same randomness, up to a controlled level. 
Examples of designs include random computational basis states and random stabilizer states on qubits, which form exact $1$- and $3$-designs respectively,
and approximate $k$-designs which can be prepared efficiently with local random quantum circuits~\cite{Harrow2009,brandao2016local, schuster2025random}.
\end{tcolorbox}

\begin{figure*}
    \centering
    \includegraphics[width=\linewidth]{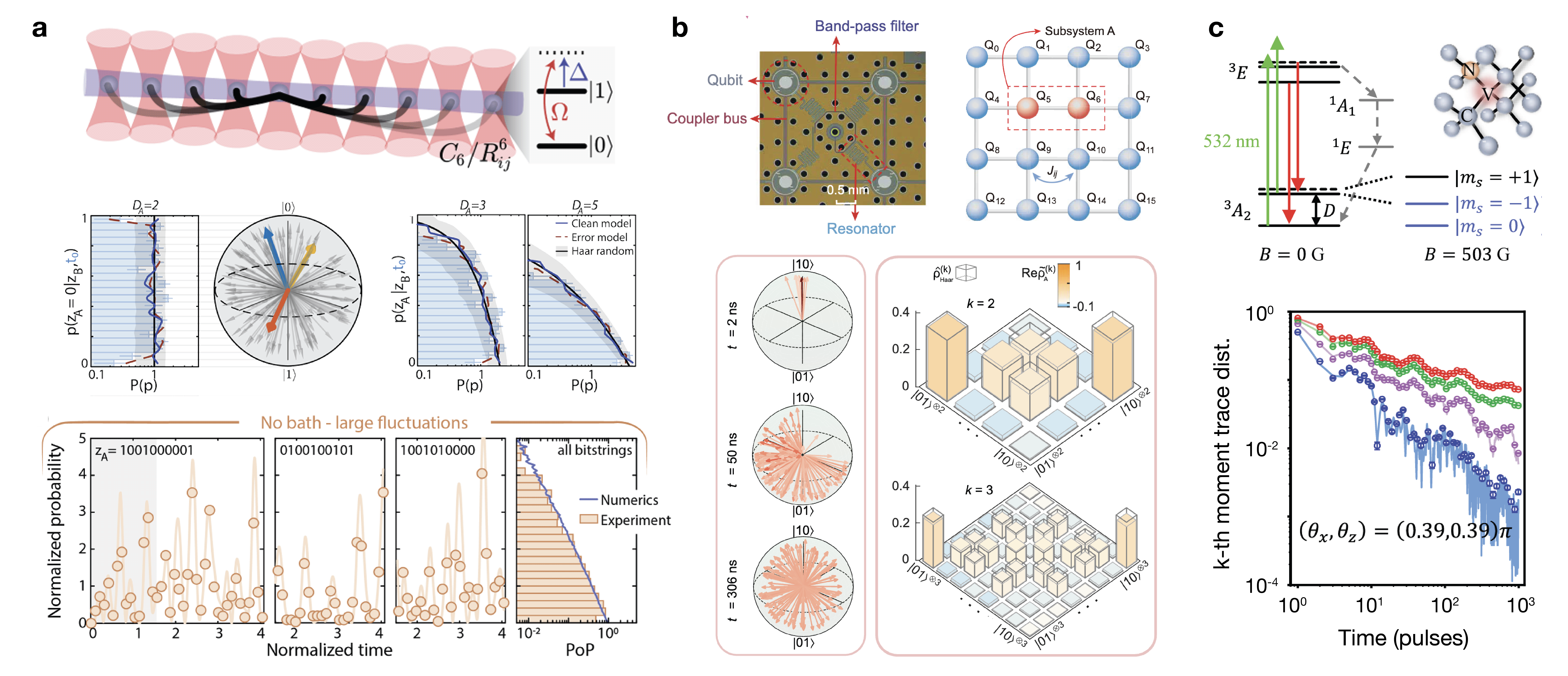}
    \caption{\textbf{Experimental evidence for deep thermalization and Hilbert space ergodicity.} A number of experiments have been performed, including on neutral atom arrays~\cite{choi2023preparing,shaw2024universal}, superconducting qubits~\cite{yan2025characterizing} and nitrogen vacancy (NV) centers~\cite{liu2026observation}. \textbf{a.}~Signatures of Haar randomness were originally observed in a Rydberg atom array (top), in which Hamiltonian quench evolution produces a state at infinite effective temperature. 
    Middle: This is revealed by the  probability-of-probabilities (PoP) of the conditional probabilities $p(z_A|z_B)$ estimated from computational-basis measurements. Bottom: HSE is also revealed by PoP statistics, here of the global probabilities $p(z)$ aggregated over time (Adapted from Refs.~\cite{choi2023preparing,shaw2024universal}). 
    \textbf{b.}~Deep thermalization was also observed in the Hamiltonian quench evolution of an array of superconducting qubits. In addition to the PoPs, the high repetition rate of superconducting qubits allow for the measurement of each projected state via state tomography: the ensemble moments $\rho_\mathcal{E}^{(k)}$ were found to converge to the Haar moments
    (Adapted from Ref.~\cite{yan2025characterizing}).
    \textbf{c.}~HSE was studied in the quantum dynamics of NV centers driven by pulse sequences. Different pulse sequences generate temporal ensembles which are $k$-designs to varying orders $k$ (Fig.~\ref{fig:venn_diagrams}), as revealed by the (non-)convergence of 
    the ensemble moments $\rho_\mathcal{E}^{(k)}$ to the Haar moments over time, as measured by tomographic single-qubit measurements (Adapted from Ref.~\cite{liu2026observation}). 
    } 
    \label{fig:expt_dat}
\end{figure*}

\noindent\textbf{
Experimental discoveries of emergent Haar randomness.
} 
Arguably the conceptually simplest incarnation of universal randomness in deep thermalization and HSE is the emergence of the {\it Haar ensemble} (Box~\red{1}), the collection of quantum states which is {\it uniformly} (and hence ``maximally randomly'') distributed over Hilbert space~\cite{mele2024interoduction}. 
Its appearance can be motivated from our understanding of standard quantum thermalization, which tells us that quantum dynamics occurring at high effective temperatures or without explicit conservation laws  lead to {\it featureless}, maximally-mixed density matrices when averaging either spatially or temporally. The Haar ensemble is the corresponding featureless quantum state ensemble, and so we may anticipate its emergence under similar physical settings.

Indeed, evidence of the Haar ensemble arising in generic quantum many-body dynamics was first provided
through far-from-equilibrium quench experiments in a Rydberg atom array quantum simulator~\cite{choi2023preparing} (Fig.~\ref{fig:expt_dat}). 
The experiments prepared high-energy product states, evolved them under the system's native Hamiltonian, then took configuration snapshots and analyzed them according to a system bipartition, as in deep thermalization, to obtain the conditional \textit{probability-of-probabilities} (PoP)---a statistics of all measurement-outcome probabilities---on a subsystem~\cite{arute2019quantum,choi2023preparing,shaw2024universal}. 
Surprisingly, even at relatively short times it was found that these statistics agree with the \textit{Porter-Thomas} (or exponential) distribution (Fig.~\ref{fig:expt_dat}a), a signature of emergent Haar-randomness~\cite{arute2019quantum,dalzell2022random,christopoulos2025universal} within the projected states.
Subsequent experiments
on Rydberg atom arrays~\cite{shaw2024universal}, superconducting qubits~\cite{yan2025characterizing}, and nitrogen vacancy (NV) centers~\cite{liu2026observation} 
(Fig.~\ref{fig:expt_dat}), have provided further evidence for the ubiquity of the Haar ensemble in settings of deep thermalization and HSE. 
In the latter two works, tomographic methods were used to explicitly construct the higher moments of the projected and temporal ensemble, respectively,  in order to  directly measure their  closeness to those of the Haar ensembles as defined by the trace distances $\Delta^{(k)}$ [Eq.~\eqref{eq:k_trace_dist}].

Theoretical studies have concomitantly supported these experimental observations and firmly established the emergence of the Haar ensemble under many settings~\cite{cotler2023emergent,ho2021exact,ippoliti2022dynamical,ippoliti2022solvable,wilming2022hightemperature,pilatowskycameo2023complete,chan2024projected,ghosh2025late}. 
For example, Ref.~\cite{cotler2023emergent} rigorously showed  that  the projected and temporal ensembles of states arising from a {\it single instance} of a deep random quantum circuit are typically uniformly distributed when the bath is large, while Ref.~\cite{ho2021exact} proved that the projected ensemble constructed from  deterministic, specially-solvable quantum dynamics called dual unitary circuits exactly converges to the Haar ensemble in the large-bath limit (see also~\cite{claeys2022emergent}).
Refs.~\cite{pilatowskycameo2023complete,pilatowskycameo2024hilbertspace,pilatowsky2025quantum,pilatowskycameo2025critically} moreover proved the emergence of the Haar ensemble within the temporal ensembles of a large class of quantum systems driven by aperiodic but deterministic time-dependent Hamiltonians $H(t)$. 

It is worth stressing that in all these findings, the appearance of the Haar random ensemble is emergent: its origins are {\it intrinsic} to the quantum system, induced either by the probabilistic nature of measurements themselves or by evolving to different times under some fixed Hamiltonian (or unitary) native to the system.  
This is conceptually distinct from the randomness arising from {\it extrinsic} sources, such as in deep quantum circuits constructed from randomly-chosen local gates~\cite{Harrow2009,brandao2016local, schuster2025random} or in the dynamics of collections of random Hamiltonians with quenched disorder~\cite{nakata2017efficient,elben2018renyi}. There the random ensemble of states generated inherits the explicit, extensive amount of classical randomness injected. 
The  important and non-trivial aspect of deep thermalization and HSE is that even ``clean,'' deterministic quantum many-body systems (for example, dynamics from a {\it single} time-independent Hamiltonian) can still spontaneously generate copious amounts of randomness, simply by virtue of the complexity they harbor.

Nevertheless, it is  an interesting practical question how the intrinsic  randomness-generating mechanism behind deep thermalization and HSE may be used in conjunction with external randomness to deliberately boost output Haar randomness, so-called \textit{randomness expansion}~\cite{mok2025optimal,ghosh2025designboostersconstanttimequantum}. 
For example, Ref.~\cite{hou2026statekdesignshamiltonianevolution} found that time-evolving random input states forming a 1-design by a fixed Hamiltonian is enough to form temporal ensembles of higher design order.

\begin{tcolorbox}[float,floatplacement=t]
{\textbf{Box 2: Scrooge ensembles and accessible information}}
\vspace{2mm}

The Scrooge ensemble was introduced and named in Ref.~\cite{jozsa1994lower}. There, they considered using ensembles of quantum states to transmit classical information: Alice draws a classical label $x$, prepares the corresponding state $|\psi_x\rangle$, and transmits it to Bob, who tries to determine $x$ by performing a measurement.
The maximum classical information that can be transmitted by a given ensemble $\mathcal{E}$ is known as its \textit{accessible information}~\cite{holevo1998capacity}:
\begin{equation}
\text{Acc}(\mathcal{E}) \equiv \text{sup}_{M} I(X;M),    
\label{eq:accessible_information}
\end{equation}
where $X$ and $M$ are random variables denoting Alice's labels and Bob's measurement outcomes, respectively. It was found that the Scrooge ensemble has the smallest accessible information over all ensembles sharing the same first moment $\rho$, i.e., is maximally ``stingy" in the information it reveals, leading to its moniker. 
The Scrooge ensemble can be sampled by a two step process:
\begin{enumerate}
    \item Sample a Haar-random state $|\phi\rangle$.
    \item Deform it into the state $\sqrt{\rho}|\phi\rangle/p_\rho(\phi)^{1/2}$ and accept with probability $p_\rho(\phi)\equiv \langle \phi|\rho|\phi\rangle$.
\end{enumerate}
In this sense, the Scrooge ensemble is a version of the Haar ensemble ``deformed" by ${\rho}$.
By construction the Scrooge ensemble satisfies a minimum information principle (minimizing accessible information); it also obeys a maximum entropy principle, in terms of an \textit{ensemble entropy}  defined on the probability distributions over Hilbert space~\cite{mark2024maximum}. 

\vspace{2mm}
\includegraphics[width=\textwidth]{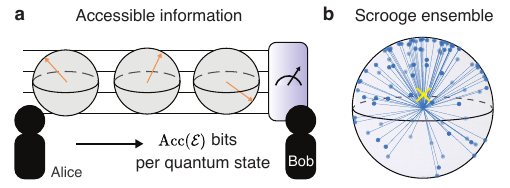}
\captionof{figure}{\textbf{Accessible information and Scrooge ensemble.} \textbf{a.} The accessible information quantifies the classical information that Alice can send to Bob by transmitting quantum states from an ensemble. \textbf{b.} The Scrooge ensemble (of a density matrix $\rho$) has the smallest accessible information over ensembles sharing the same first moment $\rho$. 
Adapted from Ref.~\cite{mark2024maximum}.}

\end{tcolorbox}

\begin{figure*}
    \centering
\includegraphics[width=\linewidth]{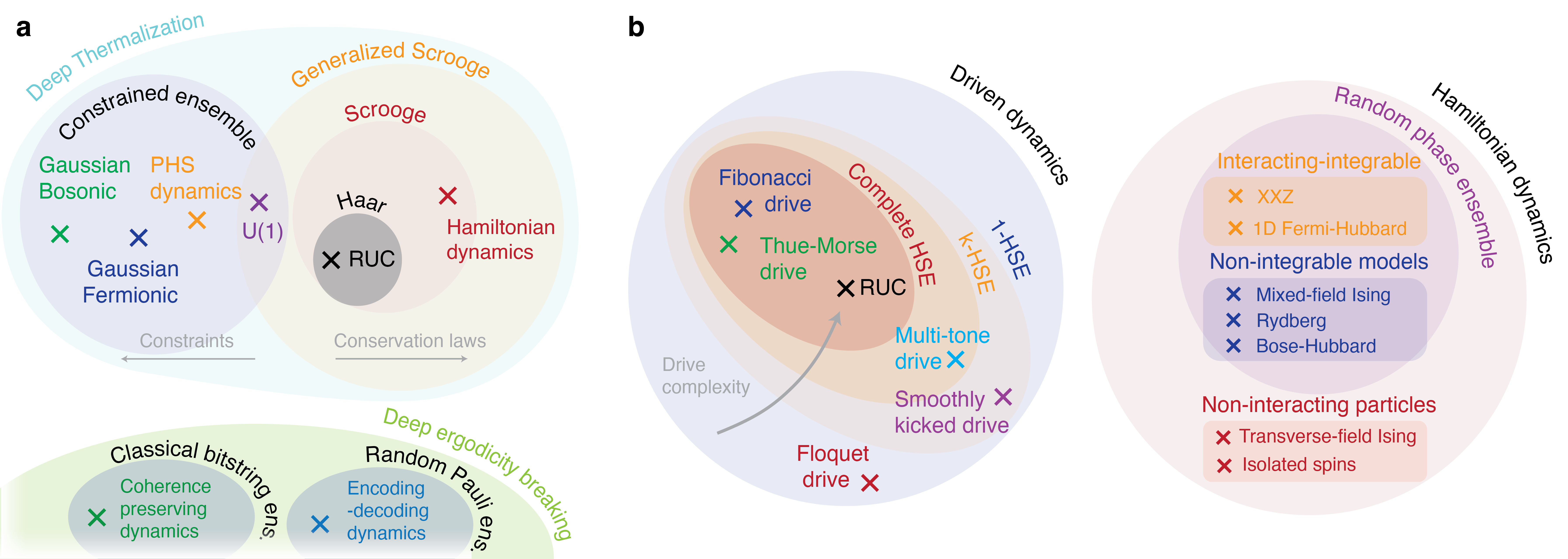}
    \caption{\textbf{Evolving landscape of deep thermalization and Hilbert space ergodicity.}  We catalog several examples of universal distributions discovered. \textbf{a.} Conservation laws or constrained dynamics result in different ensembles of states in deep thermalization, enumerated above.
    Recent work also discovers violations of deep thermalization, dubbed \textit{deep ergodicity breaking} and associated transitions between these phases.
    \textbf{b.} Hilbert space ergodicity (HSE) manifests differently in driven dynamics and in time-independent Hamiltonian dynamics. Left: in the former, systems can uniformly explore Hilbert space (``complete HSE"), or they may only explore $k$-designs (``$k$-HSE"). Complete HSE can arise either from high-complexity driven dynamics such as random unitary circuits (RUCs) or from low-complexity drives, e.g.~the Fibonacci drive. Right: In Hamiltonian dynamics, the temporal ensemble converges to the Scrooge ensemble for interacting-integrable and non-integrable models~\cite{mark2024maximum,mcginley2025scrooge,mok2026nature}, but is essentially non-ergodic in systems of non-interacting particles.}
    \label{fig:venn_diagrams}
\end{figure*}

\noindent\textbf{Beyond Haar-randomness: Scrooge ensembles and maximum-entropy principles.} While {Hilbert-space ergodicity} and {deep thermalization} originally referred to the universal emergence of the Haar ensemble, these terms have since expanded in scope to include the emergence of quantum state distributions that are ``maximally random up to constraints'' present in dynamics (Fig.~\ref{fig:venn_diagrams}). 
Constraints such as the conservation of energy are fundamental to the notion of temperature in regular thermalization itself, and consequently Haar-randomness in the projected ensemble can only be obtained at infinite temperature, when the reduced density matrix is maximally mixed $\rho_A \propto \mathbb{I}_A$~\cite{choi2023preparing,cotler2023emergent,wilming2022hightemperature,bhore2023deep}.
Similarly, the temporal ensemble cannot generically exhibit Haar-randomness if the initial state has nontrivial expectation values on conserved quantities.
Remarkably, even at finite temperatures or in the presence of conservation laws, subsequent work found that universal forms of randomness still arise, suitably modified, in the projected and temporal ensembles. These adhere to a generalized maximum entropy principle: one of \textit{minimum information revealed}, subject to prior knowledge from the constraints.

The key to understanding this constrained emergent randomness is to adopt a quantum-information-theoretic viewpoint. Specifically, one can pose two questions: (i) how does one quantify the amount of (classical) information that an ensemble can encode? And (ii) how much of such information should one generally expect in the  projected and temporal ensembles? 


The answer to question (i) is known as the \textit{accessible information} in quantum information theory~\cite{holevo1998capacity}. In particular, Ref.~\cite{jozsa1994lower} investigated the possible range of accessible information that a given ensemble $\mathcal{E}$ can have. 
An upper bound--known as the Holevo bound--is provided by the von Neumann entropy $S(\rho)$ of its mean state $\rho$, the density matrix (in the case where the ensemble states are pure), and is achieved by an ensemble of mutually orthogonal states~\cite{holevo1998capacity}.
Conversely, a lower bound is given by a quantity dubbed the \textit{subentropy} $Q(\rho)$ of $\rho$~\cite{jozsa1994lower}.
It was further shown that given a density matrix $\rho$, the quantum state ensemble attaining the lower bound is the so-called \textit{Scrooge ensemble} (Box~\red{2})--a collection of states that are maximally ``stingy'' with their information (hence the Scrooge moniker). 
Intuitively, the Scrooge ensemble can be understood as a particular deformation of the  Haar random ensemble constructed to preserve its ``maximally random'' nature whilst constrained to achieve a particular first moment $\rho$. 

Question (ii) pertains to the physical principles governing the dynamics of many interacting quantum particles.
Generally, one may anticipate that, under interacting many-body dynamics, local observables or information should be well described by a ``maximum entropy principle.'' 
Indeed, the Gibbs state---which describes steady states in conventional quantum thermalization---is obtained by maximizing the von Neumann entropy subject to energy conservation. 
In the context of the projected and temporal quantum state ensembles, one may therefore similarly expect that the information-stingy Scrooge ensembles~\cite{parfionov2006lazy,goldstein2006distribution,reimann2008typicality} should describe their emergent behavior\footnote{We note this had been anticipated in the mathematical physics literature under the name of ``Gaussian Adjusted Projected'' (GAP) ensembles~\cite{goldstein2006distribution,goldstein2016universal}.} in late-time quantum dynamics.


This expectation is borne out: it has been found that Scrooge ensembles (and refined variants, discussed next) generically emerge both in deep thermalization and HSE. 
Ref.~\cite{mark2024maximum} showed this for the temporal and projected ensembles generated by time-independent Hamiltonian dynamics under the assumption that energy eigenvalues of the Hamiltonian do not conspire to be resonant (the so-called ``$k$-th no-resonance conditions'', recently proven to generically hold in local Hamiltonians~\cite{lee2026sustained}):
the limiting ensembles are given by the Scrooge ensemble associated with the diagonal ensemble and local Gibbs states respectively. 
They have also been found in the projected ensembles of localized systems~\cite{manna2025projected}, shallow two-dimensional circuits~\cite{liu2026conditional}, the Sachdev-Ye-Kitaev model~\cite{liu2026emergence}, and in so-called dynamical ensembles~\cite{mcginley2025scrooge}. 

The maximum entropy principle also guides our understanding of deep thermalization when measurements of the bath--which generate the projected states of the projected ensemble--also reveal additional information about the subsystem. 
Indeed, some choices of measurement bases may be more informative than others, e.g., ~if the measurement basis is correlated with the conserved energy operator or a globally conserved charge. This results in the projected states being more distinguishable by such information, and consequently harboring less randomness than a standard Scrooge ensemble. 
Even so, a variant of the Scrooge ensemble, which accounts for this additional information, is predicted to emerge in these cases. Dubbed the \textit{generalized Scrooge ensemble} (GSE)~\cite{mark2024maximum}, 
this is a mixture of Scrooge ensembles which minimizes the {additional} information provided by the projected quantum state (known as the \textit{interaction information}).
Numerical studies for energy-conserving and $U(1)$ charge conserving dynamics for variable measurement bases support this prediction~\cite{mark2024maximum, chang2025deepcharge,varikuti2024unraveling}. 
A distinction between these two cases is worth mentioning.
While the former (energy-conserving) GSE seemingly requires exponentially many parameters to specify, the latter GSE ($U(1)$-charge conserving) requires merely a polynomial number. Precisely characterizing the minimum amount of information needed to specify the limiting form of the projected ensemble in deep thermalization--i.e., its {\it complexity}--is an interesting direction of future work.

Thus far, studies of deep thermalization have chiefly focused on extended quantum spin systems. A number of works have begun extending its scope to different physical settings, such as to Gaussian bosonic (or continuous-variable)~\cite{liu2024deepgaussian} and Gaussian fermionic~\cite{bejan2025matchgate, lucas2023generalized} systems (Fig.~\ref{fig:venn_diagrams}). 
Due to the non-interacting nature of dynamics, the projected states are Gaussian states, which are qualitatively distinct from the Haar random states over the Hilbert space discussed previously. Yet, maximum entropy principles have again been found to be a powerful and useful organizing principle to describe the emergent randomness in these settings: it was shown that the limiting projected ensembles of states are also maximally entropic in their appropriate  spaces~\cite{liu2024deepgaussian,lucas2023generalized,bejan2025matchgate}. In particular, \cite{liu2024deepgaussian} derived that the so-called ``Gaussian Scrooge ensemble'' emerges in Gaussian continuous-variable quantum systems. Just like the Scrooge ensemble for quantum spin systems,   the Gaussian Scrooge ensemble similarly harbors the least accessible information across all pure Gaussian state unravelings of a given density matrix. Investigating if such maximum entropy principles may further provide understanding of state ensembles arising in other physical scenarios would be an interesting avenue of future work.

\noindent\textbf{Discussion and outlook.}
Deep thermalization and Hilbert space ergodicity address fundamental questions at the heart of quantum many-body dynamics: What is the universal structure of correlations between a system and its environment? Does physical dynamics explore the Hilbert space ergodically?
Rapid experimental developments in quantum science have provided fresh insights on these foundational questions, and ideas from quantum information theory have proven instrumental in formulating answers. However, despite substantial progress, the theoretical frameworks we have described are still young: many exciting directions are yet to be explored. 

\textit{Connection with thermalization---}Deep thermalization is, in the precise sense explained earlier, a stronger notion than conventional thermalization. 
But are there more profound connections to be uncovered between the two? Recently, refinements of the eigenstate thermalization hypothesis have been proposed in order to characterize higher-order quantum fluctuations of local observables~\cite{kaneko2020characterizing,pappalardi2022eigenstate}. While deep thermalization characterizes a different type of fluctuation (over bath states), does either notion imply the other? 
More generally, under what conditions---in terms of dynamics or of eigenstates---does conventional thermalization imply deep thermalization? What is the interplay between deep thermalization and thermalization timescales set by locality~\cite{shrotriya2023nonlocality,mandal2026locality} or approximately conserved quantities~\cite{pawlik2026restricted}? 
Answering these questions would enhance our understanding of thermal equilibrium, randomness, and information hiding in nature.

\textit{Universality in more general information structures---}
Projected ensembles of pure states characterize a particular pattern of information: conditional correlations between quantum information on a local subsystem $A$ and classical data on a bath $B$. These correlations are intrinsically interesting and have also begun to be used as a tool to understand other properties of many-body systems such as magic~\cite{sarma2026universal} or the entanglement spectrum~\cite{spasicmlacak2026unravellinglihaldane}.

However, more complex patterns are also possible and natural: discarding (tracing over) some of the bath information yields projected ensembles of mixed states, which recent works have begun to study~\cite{yu2025mixed,sherry2025mixed, milekhin2025observable, mandal2026partial,logaric2025hilbert}.
Other patterns of information arise naturally in processes rather than states: there, the subsystem of interest may comprise both ``input'' and ``output'' quantum degrees of freedom, giving projected ensembles of unitary~\cite{cheng2025emergent,bejan2026projected} or more generally non-unitary~\cite{tran2023measuring,huang2026kraus} time evolution operators. Universal randomness in such objects finds applications in quantum error correction~\cite{cheng2025emergent,bejan2026projected,venkatesh2026quantumthermalization} and state learning~\cite{tran2023measuring}. 

The temporal ensemble can likewise be extended to open systems. While HSE as stated applies only to unitary evolution, a nontrivial generalization for noisy dynamics was found in the form of mixtures of Scrooge-random states~\cite{shaw2024universal}. These controlled deviations from Haar randomness in various axes means that experimental signatures such as the PoP have been found to take on successively more general, yet still universal forms such as the Erlang, hypoexponential~\cite{shaw2024universal} or Wishart~\cite{mcginley2025scrooge} distribution.

These generalized settings push the study of maximum entropy principles in new directions which, while rapidly developing, remain far less understood than their pure-state counterparts.
Interestingly, concurrent efforts to define ``intrinsically mixed'' phases of matter~\cite{sang2024mixed} point to a broader program: to characterize universality in multipartite information structures of open quantum many-body systems, across equilibrium and dynamical settings.

\textit{Phase transitions and universality classes---}One of the most fruitful research directions in quantum thermalization focuses on its exceptions: quantum ergodicity-breaking phenomena such as localization~\cite{abanin2019colloquium}, scarring~\cite{moudgalya2022quantum} and integrability~\cite{gogolin2011absence}. Does similarly rich phenomenology arise in deep thermalization? 
Recent work~\cite{liu2025coherence} provided the first instance of a ``deep-ergodicity breaking'' transition (Fig.~\ref{fig:venn_diagrams}): dynamics that always achieves conventional thermalization, yet where the projected ensemble changes abruptly from a maximally entropic form (Haar) to a   minimally entropic one (``classical bitstring'') upon tuning a parameter. This transition is explained through the lens of \textit{quantum resource theory}:
if the amount of  coherence (the resource of superposition) in the system is below a critical threshold, the projected ensemble is obstructed from being maximally entropic. Analogous thresholds emerge also for more general resources such as non-stabilizerness~\cite{feng2026quantum,loio2025quantumstatedesignsmagic,vairogs2024extracting}, and in projected ensembles of mixed states~\cite{sherry2026information}. 

These discoveries show robust exceptions to the maximum entropy principle, but what is the full landscape of such exceptions? And which physical principles, beyond quantum resource constraints, underpin them? 
Further, are there connections to the rich physics of measurement-induced phase transitions~\cite{skinner2019measurement,li2019measurement}, where sharp changes in entanglement structures in post-measurement states are similarly found? Recent findings of HSE in continuously monitored systems~\cite{wu2026exacthilbertspaceergodicitycontinuous} suggest that the two phenomena could indeed coexist. 
An overarching goal is to develop a systematic taxonomy of deep thermalization and HSE phases and phase transitions into distinct universality classes, akin to the symmetry-based classification of equilibrium phases of matter. 

\textit{Experiments, postselection, and learnability---}Deep thermalization originates as an experimental discovery, yet its experimental characterization poses significant challenges in large systems. 
All known methods to directly reconstruct the projected and temporal ensembles require a number of samples that scale exponentially with system size.
In deep thermalization, this obstacle is known as the \textit{post-selection problem} (multiple observations of the same exponentially rare measurement outcome $z_B$ are needed to learn the ensembles).
In HSE, while states can be prepared deterministically, their characterization remains expensive---both tomography and the PoP have sample complexities scaling with the Hilbert space dimension.

Fortunately, deep thermalization is not a property of individual projected states, but rather of the whole ensemble. As such, it can be probed by collective statistical properties without ever resolving individual states. Quantum-classical correlators (or ``computationally-assisted observables")~\cite{garratt2024probing,mcginley2024learning} accomplish this by correlating experimental and classically simulated measurement outcomes, offloading the complexity onto classical computation and thus enabling the study of deep thermalization up to the frontier of quantum advantage.
As quantum devices surpass this frontier, which ensemble properties remain efficiently learnable? Conversely, are there any features that are provably hard to learn? The latter may provide a concrete bridge between deep thermalization and applications in quantum cryptography~\cite{chakraborty2025fast,cui2025random}. 
A radically different direction for experiments is to avoid sampling of projected states altogether: is it possible to directly and efficiently prepare the moment states $\rho^{(k)}_\mathcal{E}$, opening up new approaches to study quantum statistical mechanics? 
Recent work~\cite{du2026cloning} has begun exploring the  resources needed to characterize projected ensembles--namely their computational or sample complexity and their fundamental trade-offs.

\textit{Applications---}Beyond fundamental science, the emergent randomness from deep thermalization and HSE has immediate applications in quantum information science. Quantum randomness is an essential ingredient in a wide range of protocols, from device benchmarking~\cite{arute2019quantum} to state learning and certification~\cite{elben2023randomized,huang2025certifying}, quantum machine learning~\cite{tran2026universality}, and beyond; however, implementing such randomness typically requires gate-level programmability of the hardware. Deep thermalization and HSE offer ways to produce this randomness even in physical settings with more limited control, thus porting a rich toolkit of randomized protocols to analog simulators~\cite{choi2023preparing,mark2022benchmarking,zhang2023superconducting,shaw2023benchmarking,andersen2025thermalization,tran2023measuring}.
As this program continues to develop, an outstanding question is whether Scrooge randomness---which emerges organically from deep thermalization and HSE, but is more challenging to implement on digital hardware---could unlock unique applications beyond Haar-randomness. 

Ultimately, the developments of deep thermalization and Hilbert space ergodicity reveal that beneath the thermal averages of textbook statistical mechanics lies further universal structure, only now coming into focus through the combined lenses of quantum information theory and modern quantum experiments. The most exciting question for the future, then, is a simple one: what else is hiding in the depths?

\vspace{0.25cm}
\noindent\textbf{Acknowledgements}\newline
{We thank the following for insightful discussions and prior collaborations: Joonhee Choi, Jordan Cotler, Andreas Elben, Xiaozhou Feng,  Tobias Haug, Hsin-Yuan Huang, Qi Camm Huang, Pavel Kos, Chang Liu, Wenquan Liu, Tudor Manole, Max McGinley, Dariel Mok, Zou-wei Pan, Saúl Pilatowsky-Cameo, John Preskill, Gil Refael, Thomas Schuster, Adam L.~Shaw, Federica Surace,  and Xie-hang Yu.}

We acknowledge support by the NSF QLCI Award OMA-2016245, the DOE (DE-SC0021951), the Institute for Quantum Information and Matter, an NSF Physics Frontiers Center (NSF Grants PHY-1733907, PHY-2317110),
the Center for Ultracold Atoms, an NSF Physics Frontiers Center (NSF Grants PHY-1734011 and PHY-2317134), Army Research Office MURI program (W911NF2010136), NSF CAREER award 2237244, the U.S. Department of Energy, Office of Science, Office of Advanced Scientific Computing Research under Award Number DE-SC0025615, the Singapore National Research Foundation (NRF) Fellowship NRF-NRFF15-2023-0008 and through the National Quantum Office, hosted in A*STAR, under its Centre for Quantum Technologies Funding Initiative (S24Q2d0009), and the
Heising-Simons Foundation (grant \#2024-4851) and the
Alfred P. Sloan Fellowship.
Support is also acknowledged from the U.S. Department of Energy, Office of Science, National Quantum Information Science Research Centers, Quantum Systems Accelerator.



\stoptoc

 \bibliography{main}

\end{document}